\documentclass[aps,pra,twocolumn,showpacs,groupedaddress]{revtex4-1}
\usepackage[latin1]{inputenc} 
\usepackage[T1]{fontenc} 
\usepackage[pdftex]{graphicx}					
\usepackage{amsmath} 							
\usepackage{amssymb}							
\usepackage[colorlinks=false]{hyperref}			

\renewcommand{\b}[1]{\mathbf{ #1}}									
\DeclareMathOperator{\sech}{sech}

\begin{document}

\title{Quantum measurement-induced dynamics of many-body ultracold bosonic and fermionic systems in optical lattices}

\author{Gabriel Mazzucchi}
\email{gabriel.mazzucchi@physics.ox.ac.uk}
\author{Wojciech Kozlowski}
\email{wojciech.kozlowski@physics.ox.ac.uk}
\author{Santiago F. Caballero-Benitez}
\author{Thomas J. Elliott}
\author{Igor B. Mekhov}
\affiliation{Department of Physics, Clarendon Laboratory, University of Oxford, Parks Road, Oxford OX1 3PU, United Kingdom}
\date{\today}

\begin{abstract}
Trapping ultracold atoms in optical lattices enabled numerous breakthroughs uniting several disciplines. Coupling these systems to quantized light leads to a plethora of new phenomena and has opened up a new field of study. Here we introduce a physically novel source of competition in a many-body strongly correlated system: We prove that quantum backaction of  global measurement is able to efficiently compete with intrinsic short-range dynamics of an atomic system. The competition becomes possible due to the ability to change the spatial profile of a global measurement at a microscopic scale comparable to the lattice period without the need of single site addressing. In coherence with a general physical concept, where new competitions typically lead to new phenomena, we demonstrate novel nontrivial dynamical effects such as large-scale multimode oscillations, long-range entanglement and correlated tunneling, as well as selective suppression and enhancement of dynamical processes beyond the projective limit of the quantum Zeno effect. We demonstrate both the break-up and protection of strongly interacting fermion pairs by measurement. Such a quantum optical approach introduces into many-body physics novel processes, objects, and methods of quantum engineering, including the design of many-body entangled environments for open systems.
\end{abstract}
\maketitle
\section{Introduction}

\begin{figure}[b]
\includegraphics[width=0.4\textwidth]{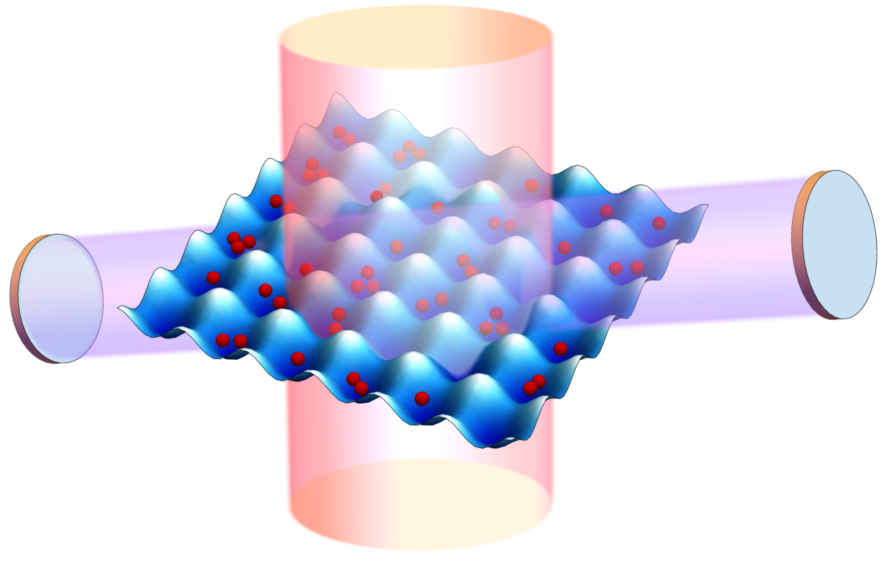}
\caption{\label{setup} Setup. Atoms in an optical lattice are probed by a coherent light beam, and the light scattered at a particular angle is enhanced and collected by a leaky cavity. The photons escaping the cavity are detected, perturbing the atomic evolution via measurement backaction. This process competes with the usual (Bose-)Hubbard dynamics and leads to the novel phenomena described in the text.}
\end{figure}

Ultracold gases trapped in optical lattices created by classical laser beams represent a successful interdisciplinary field: Atomic systems allow quantum simulations of phenomena predicted in condensed matter and particle physics, and find applications in quantum information processing~\cite{Lewenstein}. Coupling quantum gases to quantized light (cf. Refs. \cite{Mekhov2012,ritsch2013} and references therein) broadens the field even further with opportunities to obtain novel quantum phases and light-matter entanglement \cite{Moore1999, Chen2009, Caballero2015, Caballero2015a}. Here we propose how the phenomenon of quantum measurement backaction can be made efficiently compete with standard short-range processes in optical lattices (tunneling and on-site interaction), with resulting novel nontrivial effects, which do not require additional quantum control \cite{Pedersen2014}. This fits well into a general physical concept, where a novel competition in many-body systems typically leads to new phenomenon. The collapse of the atomic wave function by detecting scattered light reflects one of the most fundamental manifestations of quantum mechanics \cite{HarocheBook} - measurement backaction - here due to light-matter entanglement. We go beyond recent quantum nondemolition (QND) approaches~\cite{MekhovPRL2007, Mekhov2007b, Roscilde2009, Rogers2014, Eckert2007, HaukePRA2013, MekhovPRL2009, Rybarczyk2015, Atoms}, where either many-body dynamics or measurement backaction did not play any role. We demonstrate that the key mechanism enabling us to construct efficient competition between the global backaction and short-range processes is the possibility to spatially structure the measurement at a microscopic scale comparable to the lattice period without the need for single site resolution.  By carefully choosing the geometry of light we can select which subspace of atomic dynamics is affected by measurement.  We consider a flexible setup where the global light scattering can be engineered, allowing us to suppress or enhance specific dynamical processes, realizing a spatially nonlocal quantum Zeno effect. The measurement process generates spatial modes \cite{Elliott2015} of matter fields that can be considered as designed systems and reservoirs, opening the possibility of controlling dissipations in ultracold atomic systems, without resorting to atom losses and collisions which are difficult to manipulate. The continuous measurement of the light field introduces a controllable decoherence channel into the many-body dynamics. Global light scattering from multiple lattice sites creates nontrival, spatially nonlocal coupling to the environment, which is impossible to obtain with local interactions~\cite{Daley2014,Diehl2008,RempeScience2008,Ashida2015}.  Such a quantum optical approach can broaden the field even further, allowing quantum simulation of models unobtainable using classical light, and the design of novel systems beyond condensed matter analogs. For example, both designed systems and reservoirs are represented by many-body strongly correlated systems with internal long-range entanglement. This will raise novel type of questions and stimulate further research on dynamics and states in physics of open systems and quantum engineering. Additionally, many-body processes in dissipative systems have recently attracted attention in biological systems~\cite{DieterPRB2013}.

In contrast to other approaches of studying ground or steady state properties, here we focus on the dynamics of single quantum trajectories and find effects that are not visible in the ground and steady states. For bosons, the measurement induces dynamical macroscopic superpositions (multimode Schr{\"o}dinger cat states). The resulting states have both oscillating density-density correlations with nontrivial spatial periods and long-range coherence, thus having properties of the supersolid state~\cite{EsslingerNat2010}, but in the essentially dynamical version. For strongly interacting fermions, we demonstrate how to destroy and protect fermion pairs, clearly showing the interplay between many-body dynamics of fermions and the quantum measurement backaction. Even in the strong measurement regime, where the backaction dominates the system evolution and projects it onto a single eigenstate, we find that long-range correlated tunneling emerges, leading to dynamical generation of long-range entanglement between distant sites (which can exist even in one-dimensional systems). The setup we consider is analogous to the famous cavity QED experiments~\cite{HarocheBook}, where the quantum light states in a cavity were probed and projected by measuring atoms. Here, light and matter are reversed and we probe matter-fields with light. A striking advancement is that the number of quantum matter waves (sites with trapped atoms) can be easily scaled from few to thousands by changing the number of illuminated sites, while scaling cavities is a challenge~\cite{HarocheBook}. This work, together with recent experiments where our predictions can be tested (Bose-Einstein Condensates trapped inside a cavity~\cite{EsslingerNat2010,HemmerichScience2012,ZimmermannPRL2014}, and recently a lattice in a cavity \cite{Landig2015a, Klinder2015}), will help to develop a deeper understanding of the effect of quantum measurement on the dynamics of strongly correlated many-body systems and uncover the constructive and active role of global backaction in systems with short-range interactions, thus closing the gap between quantum optics and quantum gases.

\section{Model for spatially structured global measurement}
We consider off-resonant light scattering from $N$ atoms trapped in an optical lattice with period $d$ and $L$ sites~\cite{Mekhov2012} (see Appendix A and Fig.~\ref{setup}). The light scattered at a particular angle can be selected and enhanced by a cavity~\cite{Bux2013,Kessler2014,Landig2015} with decay rate $\kappa$. Similar to classical optics, the light amplitude is given by a sum of scatterings from all atoms with coefficients dependent on their positions: $a = C (\hat{D} + \hat{B})$, where $a$ is the photon annihilation operator, $C$ is the Rayleigh scattering coefficient (see Appendix A) and  
\begin{align}\label{op}
\hat{D}=\sum_{j=1}^L J_{jj} \hat{n}_j, \quad \hat{B}=\sum_{\langle i,j \rangle}^L J_{ij} b_i^{\dagger} b_j, 
\end{align}
where $b_j$ and $\hat{n}_j=b_j^{\dagger} b_j$ are the atomic annihilation and number operators at site $j$ (${\langle i,j \rangle}$ sums over neighboring sites) and $J_{ij}$ are given by
\begin{align}
J_{ij}=\int \! w (\b{r} -\b{r}_i) u_{\mathrm{out}}^*(\b{r})u_{\mathrm{in}}(\b{r}) w (\b{r} -\b{r}_j) \, \mathrm{d}  \b{r},
\end{align}
where $w (\b{r})$ are the localized Wannier functions and $u_{\mathrm{in,out}}(\b{r})$ are the mode functions of incoming and outgoing light respectively (e.g., $u_l(\b{r})=\exp({i\b{k}_l\b{\cdot} \b{r}})$ for traveling and $u_l(\b{r})=\cos({\b{k}_l\b{\cdot} \b{r}})$ for standing waves with wave vectors $\b{k}_l$). 

In Eq.~(\ref{op}) $\hat{D}$ describes scattering from the on-site densities, while $\hat{B}$ that from the inter-site coherence terms~\cite{Kozlowski}. For well-localized atoms, the second term is usually neglected, and $a=C \hat{D}$ with $J_{jj}=u_{\mathrm{out}}^*(\b{r}_j) u_{\mathrm{in}}(\b{r}_j)$. For spin-$\frac{1}{2}$ fermions we use two light polarizations $a_{x,y}$ that couple differently to two spin densities $\hat{n}_{\uparrow j}$, $\hat{n}_{\downarrow j}$ allowing measurement of their linear combinations, e.g.,  $a_x=C \hat{D}_x=C\sum_{j=1}^L J_{jj} \hat{\rho}_{j}$ and $a_y=C \hat{D}_y=C\sum_{j=1}^L J_{jj} \hat{m}_j$, where $\hat{\rho}_{j}=\hat{n}_{\uparrow j}+\hat{n}_{\downarrow j} $ and $ \hat{m}_j=\hat{n}_{\uparrow j}-\hat{n}_{\downarrow j}$ are the mean density and magnetisation. This property has recently been used to investigate spin-spin correlations in Fermi gases \cite{Meineke2012, Sanner2012}.

We focus on a single run of a continuous measurement experiment using the quantum trajectories technique~\cite{MeasurementControl} (see Appendix B). The evolution is determined by a stochastic process described by quantum jumps (the jump operator $c=\sqrt{2 \kappa} a$  is applied to the state when a photodetection occurs) and non-Hermitian evolution with the Hamiltonian 
\begin{equation}\label{Heff}
\hat{H}_{\mathrm{eff}}=\hat{H}_0- i \hbar c^\dagger c /2
\end{equation}
between jumps, where $H_0$ is the usual (Bose-)Hubbard Hamiltonian. Importantly, the measurement introduces a new energy and time scale $\gamma=|C|^2 \kappa$, which competes with the two other standard scales responsible for unitary dynamics of closed systems (tunneling $J$ and on-site interaction $U$). If atoms scatter light independently, independent jump operators $c_j$ would be applied to each site, projecting the atomic system to a state, where the long-range coherence degrades~\cite{PichlerDaley2010}. This is a typical scenario for spontaneous emission~\cite{PichlerDaley2010,Sarkar2014}, or rather analogous local~\cite{Daley2014,Bernier2014,Hofstetter2014,Hartmann2012,RempeScience2008,Wade2015} and fixed-range~\cite{LesanovskyPRL2012,LesanovskyPRB2014} addressing and interactions. Additionally, if the light is scattered without a cavity, i.e.~in uncontrolled directions, we lose the ability to choose the measurement operator and the jump operator applied would depend on the direction of the detected photon. In contrast, here we consider global coherent scattering, where the single global jump operator $c$ is given by the sum over all sites, and the local coefficients $J_{jj}$ (\ref{op}) responsible for the atom-environment coupling (via the light mode $a$) can be engineered by optical geometry. Thus, atoms are coupled to the environment globally, and atoms that scatter light with the same phase are indistinguishable to light scattering (i.e. there is no ``which-path information''). As a striking consequence, the long-range quantum superpositions are strongly preserved in the final projected states, and the system splits into several spatial modes \cite{Elliott2015}, where all atoms belonging to the same mode are indistinguishable, while being distinguishable from atoms belonging to different modes. 

We engineer the atom-environment coupling coefficients $J_{jj}$ using standing or traveling waves at different angles to the lattice. A key mechanism, which will allow us to construct the effective competition between global measurement and local processes, is the ability to modify these couplings at very short microscopic distances. This is in striking contrast to typical scenarios of Dicke and Lipkin-Meshkov-Glick models~\cite{ParkinsPRL2008}, where the coupling is global, but rather homogeneous in space. If both probe and scattered light are standing waves crossed at such angles to the lattice that projection $\b{k}_\mathrm{in} \b{\cdot} \b{r}$ is equal to $\b{k}_\mathrm{out}\b{\cdot} \b{r}$ and shifted such that all even sites are positioned at the nodes (do not scatter light), one gets $J_{jj}=1$ for odd and $J_{jj}=0$ for even sites. Thus we measure the number of atoms at odd sites only (the jump operator is proportional to $\hat{N}_\text{odd}$), introducing two modes, which scatter light differently: odd and even sites. The coefficients $J_{jj}=(-1)^j$ are designed by crossing light waves at $90^\circ$ such that atoms at neighboring sites scatter light with $\pi$ phase difference~\cite{MekhovPRL2007,MekhovPRL2009,mekhovLP2009,mekhovLP2010,mekhovLP2011}, giving $\hat{D}=\hat{N}_{\mathrm{even}}-\hat{N}_{\mathrm{odd}}$, introducing the same modes, but with different coherence between them. Moreover, using travelling waves crossed at the angle such that each $R$-th site is indistinguishable ($(\b{k}_\mathrm{in} - \b{k}_\mathrm{out})\b{\cdot} \b{r}_j=2\pi j/R$), introduces $R$ modes with macroscopic atom numbers $\hat{N}_l$: $\hat{D}=\sum_{l=1}^R \hat{N}_l e^{i2\pi l/R}$~\cite{Elliott2015}. Here two (odd- and even-site modes) appear for $R=2$. Therefore, we reduce the jump (measurement) operator from being a sum of numerous microscopic contributions from individual sites to the sum of smaller number of macroscopically occupied modes with a very nontrivial spatial overlap between them. 

Our results are applicable in any number of dimensions and for an arbitrary system size, but for simplicity and clarity our results are often presented in one dimension. The dimensionality of the system affects key properties of the ground state such as quantum critical points and decay of atomic correlations. However, in this paper the ground state represents only a realistic initial condition and the long-range correlations are dominated by the effect of the nonlocal nature of the measurement. Therefore, the results that we will present do not strictly depend on the number of spatial dimensions but on the geometry and symmetry of the measurement scheme. The only exception is represented by the behavior of 1D fermions: This differs from any other dimension since atoms with the same spin can not pass each other.

In the following, we will show how globally designed measurement backaction introduces spatially long-range interactions of the modes, and demonstrate novel effects resulting from the competition of mode dynamics with standard local processes in a many-body system.

\begin{figure}[t!]
\includegraphics[width=0.5\textwidth]{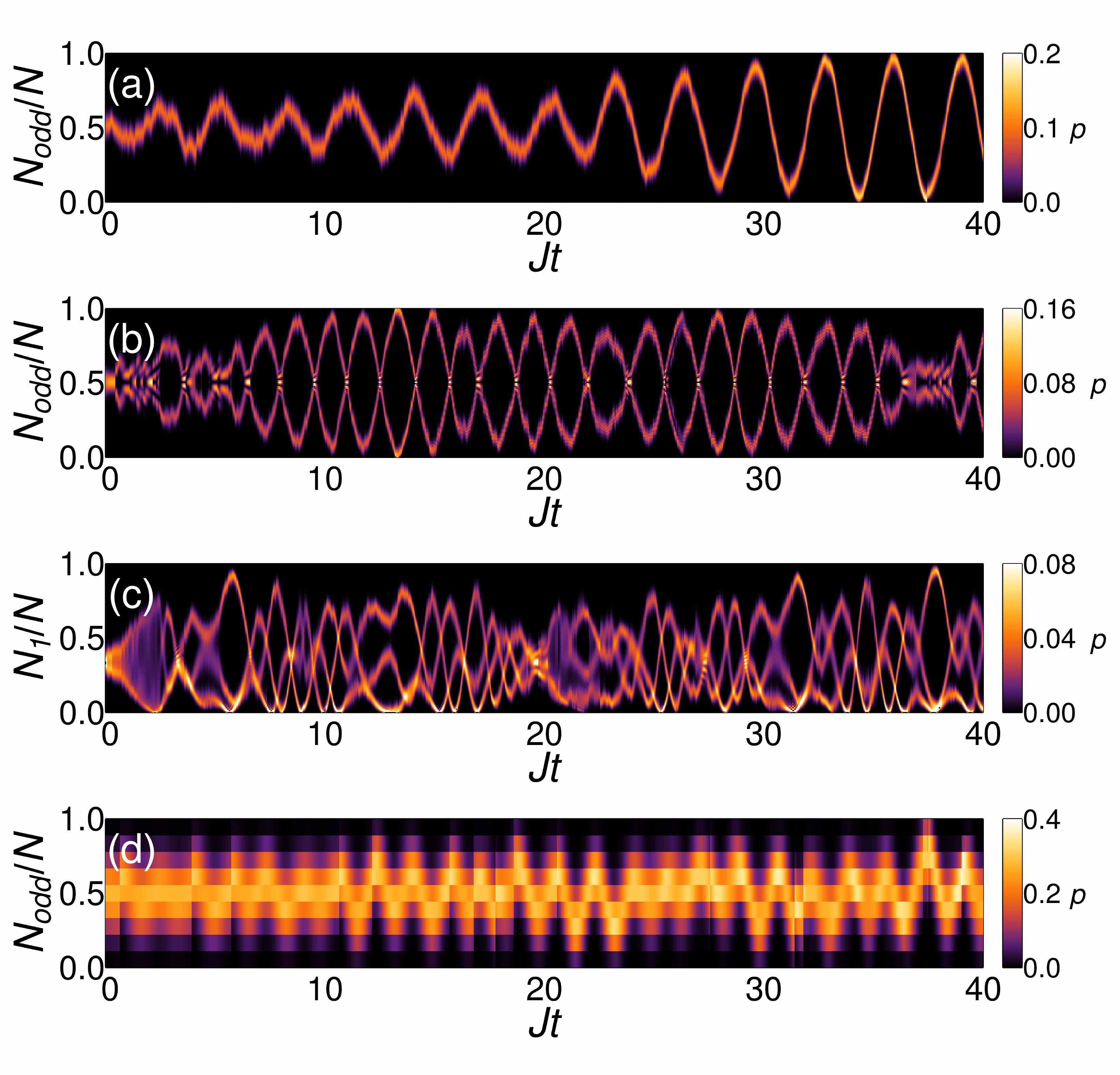}
\caption{Large oscillations between the measurement-induced spatial modes resulting from the competition between tunneling and weak-measurement backaction. The plots show single quantum trajectories. (a)-(d), Atom number distributions $p(N_l)$ in one of the modes, which show various number of well-squeezed components, reflecting the creation of macroscopic superposition states depending on the measurement configuration ($U/J=0$, $\gamma/J=0.01$, $L=N$, initial states: superfluid for bosons, Fermi sea for fermions). (a), Measurement of the atom number at odd sites $\hat{N}_\text{odd}$ creates one strongly oscillating component in $p(N_\text{odd})$ ($N=100$ bosons, $J_{jj} = 1$ if $j$ is odd and 0 otherwise). (b), Measurement of $(\hat{N}_\text{odd}-\hat{N}_\text{even})^2$ introduces $R=2$ modes and preserves the superposition of positive and negative atom number differences in $p(N_\text{odd})$ ($N=100$ bosons, $J_{jj}=(-1)^{j+1}$). (c), Measurement for $R=3$ modes (see text) preserves three components in $p(N_1)$ ($N=108$ bosons, $J_{jj}=e^{i j 2 \pi /3}$).  (d), Measurement of $\hat{N}_\text{odd}$ for fermions leads to oscillations in $p(N_\text{odd})$, though not as clearly defined as for bosons because of Pauli blocking ($L=8$ sites, $N_{\uparrow}=N_{\downarrow}=4$ fermions, $J_{jj} = 1$ if $j$ is odd and 0 otherwise). Simulations for 1D lattice.} \label{panel1}
\end{figure}

\section{Large-scale dynamics due to weak measurement}
We start with non-interacting bosons ($U/J=0$), and demonstrate that the competition between tunneling and global measurement strongly affects the dynamics of the atomic system. The weak measurement ($\gamma \ll J$) is unable to freeze the atom numbers projecting the atomic state and quantum Zeno dynamics~\cite{Facchi2009, Raimond2010, Raimond2012, schafer2014, Signoles2014} cannot be established. In contrast, the measurement leads to giant oscillations of particle number between the modes. Figures~\ref{panel1}a-c illustrate the atom number distributions in one of the modes for $R=2$ ($N_\text{odd}$) and $R=3$ ($N_1$). Without continuous monitoring, these distributions would spread out significantly and oscillate with an amplitude proportional to the the initial imbalance, i.e.~tiny oscillations for a tiny initial imbalance. In contrast, here we observe (i) full exchange of atoms between the modes independent of the initial state and even in absence of initial imbalance, (ii) the distributions consist of a small number of well-defined components, and (iii) these components are squeezed even by weak measurement.
Depending on the quantities addressed by the measurement, the state of the system has a multi-component structure which is a consequence of the photon number (intensity) $a^\dag a$ not being sensitive to the light phase. In other words, the measurement does not distinguish between all permutations of mode occupations that scatter light with the same intensity. The number of components of the atomic state, i.e.~the degeneracy of $a^\dag a$, can be computed from the eigenvalues of  $\hat{D}=\sum_{l=1}^R \hat{N}_l e^{i2\pi l/R}$ noting that they can be represented as the sum of vectors on the complex plane with phases that are integer multiples of $2\pi /R$: $N_1e^{i2\pi/R},N_2e^{i4\pi/R}, \dots  N_R $. Since the sum of these vectors is invariant under rotations by $2\pi l /R, \, \, l\in \mathbb{Z}$ and reflection in the real axis, the state of the system is 2-fold degenerate for $R=2$ and $2R$-fold degenerate for $R>2$.
Figure~\ref{panel1}b shows the superposition of two states with positive and negative $N_\text{odd}-N_\text{even}$, while Fig.~\ref{panel1}c illustrates the superposition in the three-mode case. 

One can get a physical insight into the origin of oscillations by constructing the following model. For macroscopically occupied modes of non-interacting atoms initially in the superfluid state, the $R$-modes problem can be treated analytically reducing it to $R$ effective sites. In particular, for $R=2$ (effective double-well \cite{Corney1997,Julia-Diaz2012}), we can write the atomic state as 
\begin{align}\label{wf}
|\psi\rangle =\sum_{l=0}^{N} q_l |l,N-l\rangle.
\end{align}
where the ket $|l,N-l\rangle$ represents a superfluid with $l$  atoms in the first and $N-l$ atoms in the second spatial mode. In the limit $N \gg 1$  we can describe the evolution of the system using continuous variables~\cite{Julia-Diaz2012} and define the wave function $\psi(x=l/N)=\sqrt{N} q_l$. Introducing the relative population imbalance between the two wells $z=2 x -1$ and starting from the superfluid state, the solution of the Schr\"odinger equation with Hamiltonian $\hat{H}_{\mathrm{eff}}$ (\ref{Heff}) is
\begin{align}
\psi(z,t)\propto \exp \left[{i a(t)+\frac{i z c(t)}{2 b^2(t)} +\frac{i z^2 \phi(t)}{2 b^2(t)}-\frac{(z-z_0(t))^2}{2  b^2(t)}}\right],
\end{align}
where $b(t)$ is the width of atomic distribution, $z_0(t)$ is the average population imbalance and $c(t)$ and $\phi(t)$ are real phases while $a(t)$ is complex and takes into account the decay of norm of the wave function due to the non-Hermitian term present in $\hat{H}_{\mathrm{eff}}$. Expanding the Hamiltonian in powers of $1/N$ up to second order we obtain a system of coupled differential equations for $a(t), \,b(t), \,c(t), \,z_0(t)$, and $\phi(t)$ which, in the weak measurement limit, can be linearized, leading to 
\begin{align}\label{population}
 z_0(t)=\frac{1}{2} \mathrm{e}^{-\frac{N \gamma}{2}t } \left[c(0) \sin (2 J t)+2 z_0(0) \cos ( 2 J t)\right].
\end{align}
This expression describes the evolution of the population imbalance between two quantum jumps: The atoms oscillate between the two spatial modes with decreasing amplitude and tend to restore a balanced distribution. The quantum jumps strongly affect the dynamics of the system via measurement backaction, driving the oscillations to their possible maximum amplitude, i.e.,~all the atoms oscillate between the two spatial modes. We explain this effect noting that (i) the average time between two jumps ($\sim2 / N^2 \gamma$) is much smaller than the damping time in (\ref{population}), (ii) the probability of a jump depends on the imbalance itself $p_\mathrm{jump}\propto (1+z_0)^2$ and (iii) each jump tends to increase the population imbalance. In order to calculate how the photodetection events affect the evolution of $z_0$, we compute the effect of average jump rate as $\delta z_0 / \delta t= p_{\mathrm{jump}}\, \Delta z_0 / \delta t $, where $\Delta z_0$ is the change in $z_0$ due to the jump itself. Solving this equation leads to
\begin{align}\label{jumps}
 z_{0,\mathrm{jumps}}(t)=-1+(1+z_0(0))\mathrm{e}^{N \gamma t}.
\end{align}
Therefore, the amplitude of oscillations of $z_0$ increase since the difference between the exponents in (\ref{population}) and (\ref{jumps}) is positive and the system leaves the stable point $z_0=\dot{z_0}=0$. Note that the dynamics described here is a feature of single trajectories only. This is in contrast to averaging over many runs, which corresponds to the master equation solution,  which masks these effects completely. This happens because the oscillation phase changes from realization to realization as is known from works involving single and multiple measurements~\cite{RuostekoskiPRA1997}.

The competition between measurement and atomic dynamics allows realization of multicomponent macroscopic superpositions (Schr{\"o}dinger cat or NOON states), which are useful in quantum metrology and information. Such multimode superpositions are a purely quantum effect. The multimode dynamics may recover the semiclassical character~\cite{Lee2014}, when the number of modes is reduced to 1. The method we propose does not require external control \cite{Pedersen2014,Ivanov2014, Ivanov2016} for preparing these states: By continuously monitoring the light intensity it is possible to determine when the splitting in the components reaches its maximum value (corresponding to the maximal macroscopicity \cite{Frowis2012}) and further oscillatory dynamics can be stopped by ramping up the lattice depth. Note that for $R>1$ spatial modes each photocount changes the phase difference between the various components of the atomic state, making it fragile to photon losses. However, the measurement setup can be modified to make these states more robust \cite{MekhovPRA2009}. In addition, the example in Fig.~\ref{panel1}(a) consists of only one component (here $N_\text{odd}$ is measured directly) and is therefore insensitive to decoherence due to photon losses.

It is interesting to note that the oscillating state we have described shows spatial periodicity in the density-density correlation function depending on the measurement configuration. For $R$ modes the spatial period is $Rd$ ($d$ is the lattice period). Moreover, the state also has long-range coherence between distant sites. Therefore, it has properties of the supersolid state~\cite{EsslingerNat2010} but with a nontrivial period, and it exists in the essentially dynamical version.

In contrast to bosons, dynamics for two modes of non-interacting fermions does not show well-defined oscillations (Fig.~\ref{panel1}d) due to Pauli exclusion. However, while the initial ground state is a product of $\uparrow$ and $\downarrow$ wave functions (Slater determinants), the measurement introduces an effective interaction between two spin components and the state becomes entangled by measurement~\cite{Mazzucchi2015}. The dynamics of a one-dimensional fermionic system strongly depends on the spatial profile of the measurement operator. For example, if only the central part of the lattice is illuminated (diffraction maximum) the atom transfer between the modes induced by the measurement is suppressed as the presence of an atom at the edges of the illuminated area completely forbids the atomic tunneling between the two modes, greatly decreasing the fluctuations of the measurement operator.

\begin{figure}[t!]
\includegraphics[width=0.5\textwidth]{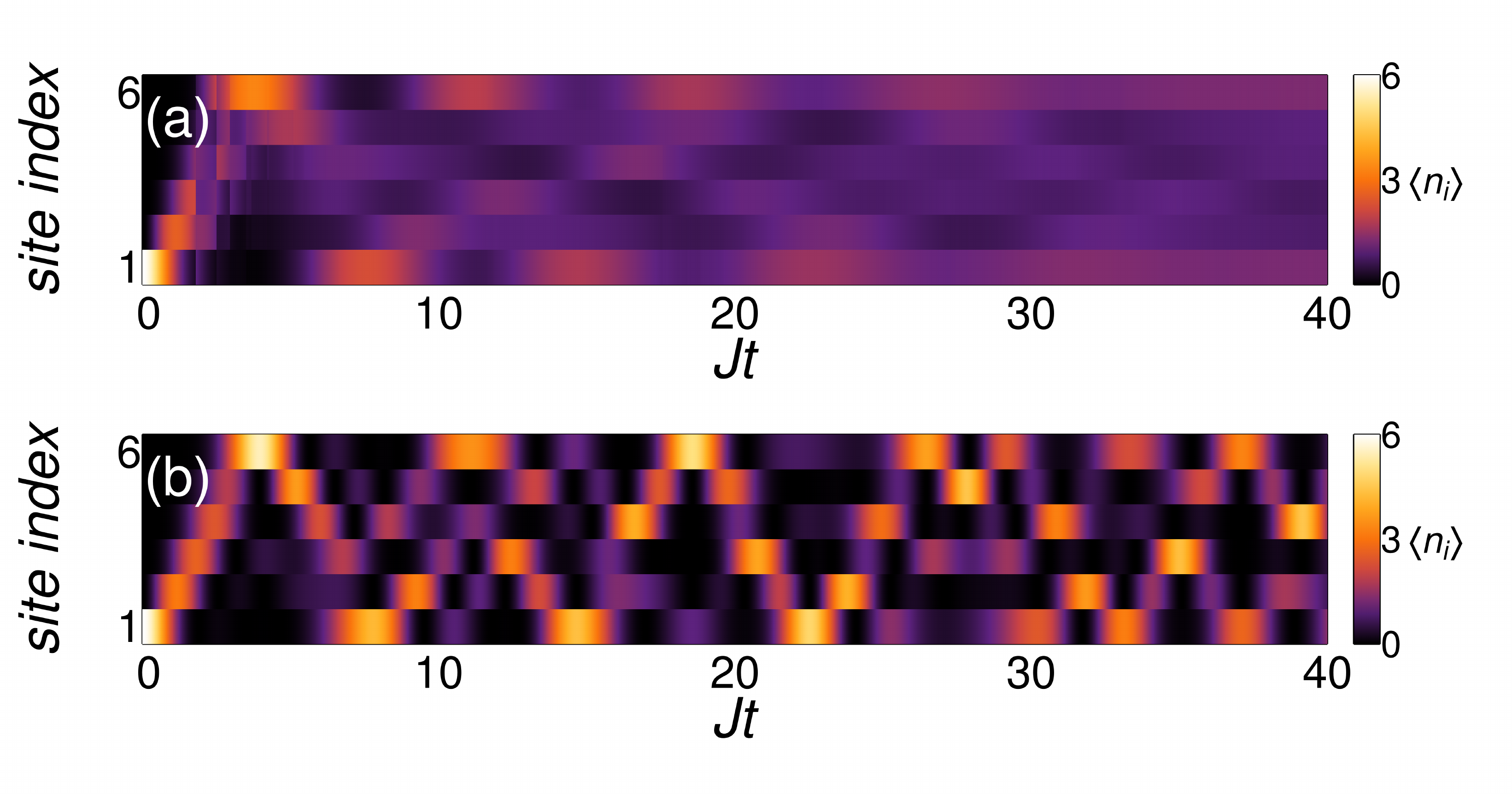}
\caption{Evolution of the on-site atomic density, while measuring the matter-field coherence between the sites $\hat{B}$. The plots show single quantum trajectories.  (a), Atoms, all initially at the edge site, are quickly spread across the whole lattice leading to the large uncertainty in the atom number, while the matter-phase related variable is defined (projected) by the measurement (bosons, $N=L=6$, $U/J=0$, $\gamma/J=0.1$, $J_{jj}=1$). (b), Atomic density spread and revival due to coherent tunnelling in the absence of measurement (bosons, $N=L=6$, $U/J=0$, $\gamma/J=0$, $J_{jj}=1$). Simulations for 1D lattice.} \label{panel1bis}
\end{figure}

Carefully choosing geometry, one can suppress the on-site contribution to light scattering and effectively concentrate light between the sites, thus \emph{in-situ} measuring the matter-field interference $b^\dag_ib_{i+1}$~\cite{Kozlowski}. In this case, $a=C\hat{B}$, and the coefficients $J_{ij}$ for $i\ne j$ from Eq. (\ref{op}) can be engineered \cite{Kozlowski}. For $J_{ij}=1$, the jump operator is proportional to the kinetic energy $\hat{E}_K=-2 \hbar J \sum_{k} b^\dagger _k b_k \cos(ka)$ and tends to freeze the system in eigenstates of the non-interacting Hamiltonian. The measurement projects to a superposition of two states with different kinetic energies: a superposition of matter waves propagating with different momenta. The measurement does not distinguish between these two states, because in a lattice, the two momentum states $|k\rangle$ and $|\pi/a-k\rangle$ interfere in the same way, but with opposite phase in between the lattice sites. The measurement freezes dynamics for any $\gamma/J$, since the jump operator and $\hat{H}_{\mathrm{eff}}$ have the same eigenstates. As a result of the detection, the atoms quickly spread across the lattice, and the density distribution becomes uncertain (Fig.~\ref{panel1bis}a), clearly illustrating the quantum uncertainty relation between the number- and phase-related variables ($\hat{n}_i$ and $b^\dag_ib_{i+1}$). Note that, in absence of measurement, such a distribution presents a periodic spread and revival due to coherent tunneling and the system does not reach a steady state (Fig.~\ref{panel1bis}b). Therefore, engineering $J_{ij}$ can lead to the measurement-based preparation of peculiar multicomponent momentum (or Bloch) states.

\section{Competition between interaction, tunneling and measurement}
As we turn on the inter-atomic interactions, $U/J\neq 0$, the atomic dynamics changes as the measurement competes with both the tunneling and this on-site interaction.  One approach is to study the ground or steady state of the system in order to map a quantum or dissipative phase diagram. This is beyond the scope of this article, because here, we adopt a quantum optical approach with the focus on the conditional dynamics of a quantum trajectory corresponding to a single experimental realization.
The resulting evolution does not necessarily reach a steady state and can occur far from the ground state of the system. 
Again, each quantum trajectory evolves differently as the detection process is determined stochastically and even states with similar expectation values of $\hat{D}$ can have minimal overlap. However, even though each trajectory is different they all have one feature in common: The uncertainty in the measured operator, $\hat{D}$, is only a function of the Hamiltonian parameters, $\gamma$, $J$, and $U$. Therefore, we average its variance over many realizations ($\langle \sigma^2_D \rangle_{\mathrm{traj}}$) as this quantity effectively describes the squeezing of the atomic distribution due to measurement. Importantly, it is not possible to access this quantity using the master equation solution: The uncertainty in the final state is very large and it completely hides any information on the spread of a single trajectory. In other words, the master equation addresses the variance of the average value of $\hat{D}$ over the trajectories ensemble ($\langle \hat{D}^2 \rangle_{\mathrm{traj}}-\langle \hat{D} \rangle_{\mathrm{traj}}^2$) and not the squeezing of a single trajectory conditioned to the measurement outcome. This again highlights the fact that interesting physics happens only at the single trajectory level. In this section we show results for the measurement of  $\hat{D}=\hat{N}_{\mathrm{odd}}$, which is robust to photon losses. Specifically, we compute the width of the atomic distribution ($\langle \sigma^2_D \rangle_{\mathrm{traj}}$) in the limit $t\rightarrow \infty$, when its value does not change significantly in time even if the atomic imbalance is not constant. Moreover, we use the ground state of the system as initial state since this is a realistic starting point and a reference for explaining the measurement-induced dynamics.

\begin{figure}
\includegraphics[width=0.5\textwidth]{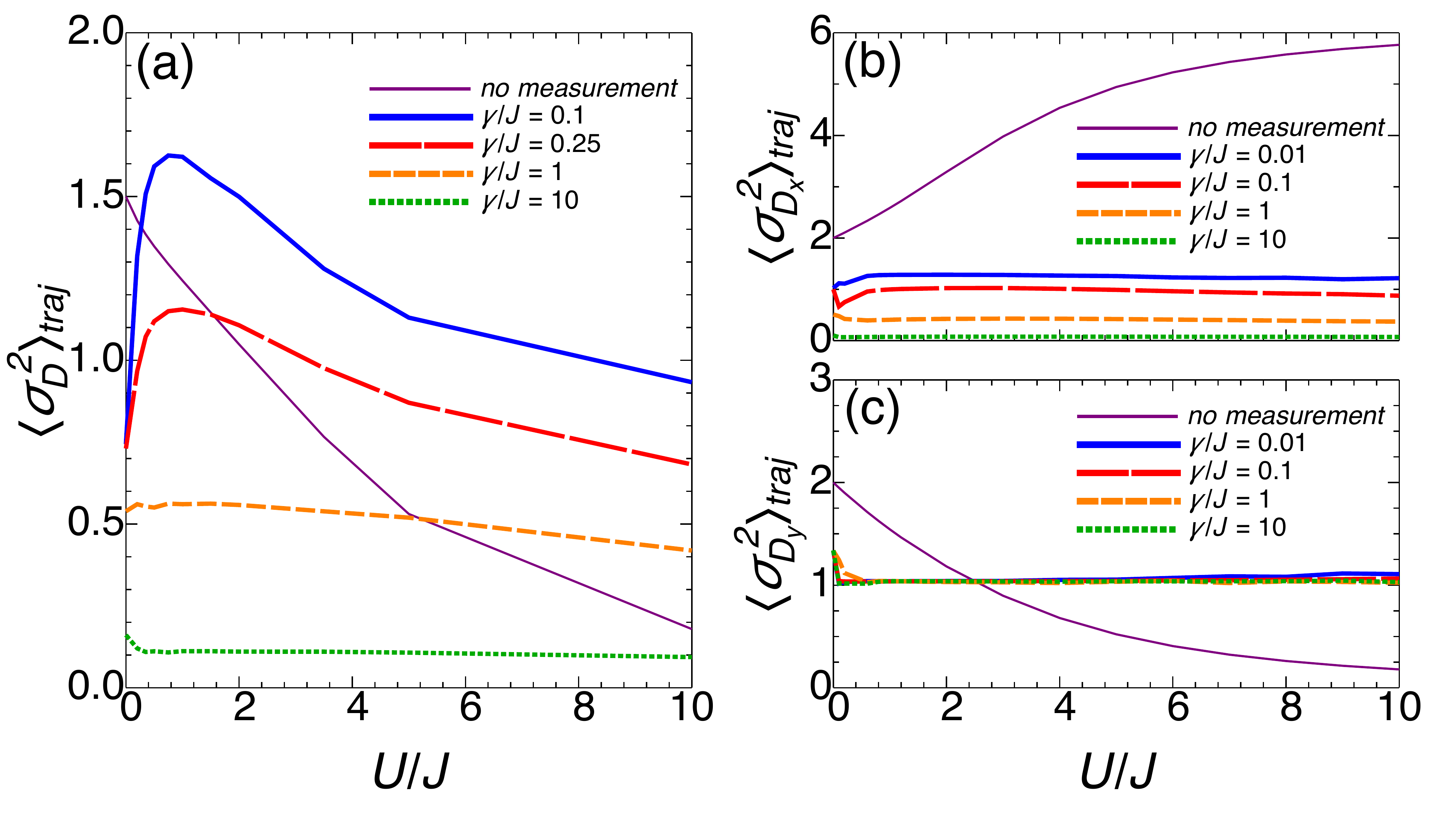}
\caption{\label{panelfluc} Atom number fluctuations demonstrating the competition of global measurement with local interaction and tunneling. Number variances are averaged over many trajectories. Error bars are too small to be shown ($\sim 1\%$), which emphasizes the fundamental nature of the squeezing.  (a), Bose-Hubbard model with repulsive interaction. The fluctuations of the atom number at odd sites $\hat{N}_\text{odd}$ in the ground state without a measurement (thin solid line) decrease as $U/J$ increases, reflecting the transition between the superfluid and Mott insulator phases. For the weak measurement, $\langle \sigma^2_D \rangle_{\mathrm{traj}}$ is squeezed below the ground state value, but then increases and reaches its maximum as the atom repulsion prevents oscillations and makes the squeezing less effective. In the strong interacting limit, the Mott insulator state is destroyed and the fluctuations are larger than in the ground state. (100 trajectories, $N=L=6$, $J_{jj} = 1$ if $j$ is odd and 0 otherwise.) (b),(c), Fermionic Hubbard model with attractive interaction; fluctuations of the total atom number at odd sites $\hat{D}_x=\hat{N}_{\uparrow \text{odd}}+ \hat{N}_{\downarrow \text{odd}}$ (\textbf{b}) and of the magnetization at odd sites $\hat{D}_y=\hat{M}_\mathrm{odd}=\hat{N}_{\uparrow \text{odd}}-\hat{N}_{\downarrow \text{odd}}$ (c). Without measurement, the interaction favors formation of doubly occupied sites so the density fluctuations in the ground state are increasing while the magnetization ones are decreasing. The measurement creates singly occupied sites decreasing the density fluctuations and increasing the magnetization ones, which manifests the break-up of fermion pairs by measurement. The measurement-based protection of fermion pairs is shown in the next figure. (100 trajectories, $L=8$, $N_{\uparrow}=N_{\downarrow}=4$, $J_{jj} = 1$ if $j$ is odd and 0 otherwise.) Simulations for 1D lattice.} 
\end{figure}

\begin{figure}
\includegraphics[width=0.5\textwidth]{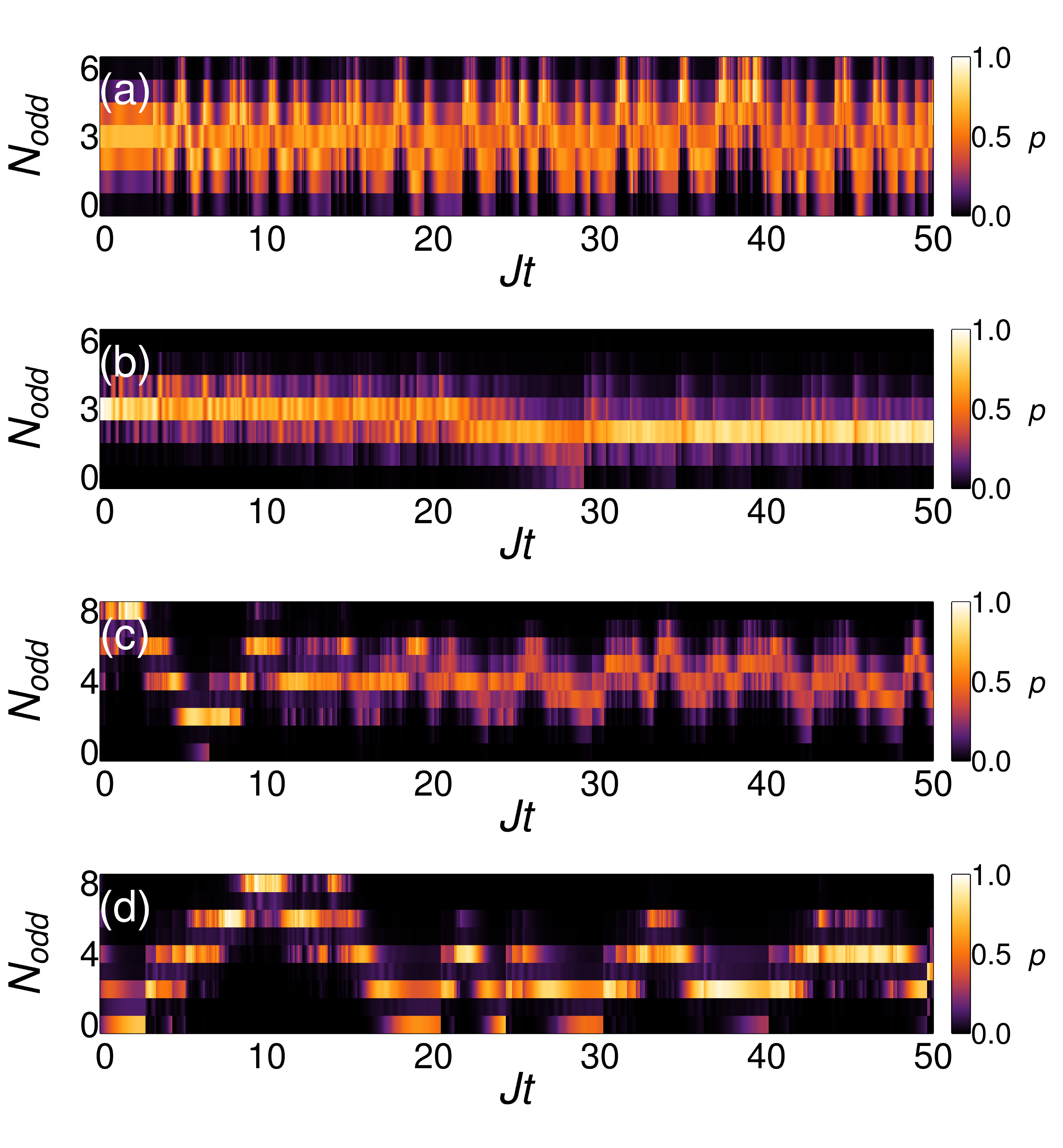}
\caption{\label{panel2} Conditional dynamics of the atom-number distributions at odd sites illustrating competition of the global measurement with local interaction and tunneling (single quantum trajectories, initial states are the ground states). (a), Weakly interacting bosons: the on-site repulsion prevents the formation of well-defined oscillations in the population of the mode. As states with different imbalance evolve with different frequencies, the squeezing due to the measurement is not as efficient as one observed in the non-interacting case ($N=L=6$, $U/J=1$, $\gamma/J=0.1$, $J_{jj} = 1$ if $j$ is odd and 0 otherwise). (b), Strongly interacting bosons: oscillations are completely suppressed and the number of atoms in the mode is rather well-defined, although less squeezed than in the Mott insulator. ($N=L=6$, $U/J=10$, $\gamma/J=0.1$, $J_{jj} = 1$ if $j$ is odd and 0 otherwise).(c), Attractive fermionic Hubbard model in the strong interaction limit. Measuring only the total population at odd sites $\hat{D}_x=\hat{N}_{\uparrow \text{odd}}+ \hat{N}_{\downarrow \text{odd}}$ quickly creates singly occupied sites, demonstrating measurement-induced break-up of fermion pairs ($L=8$, $N_{\uparrow}=N_{\downarrow}=4$, $U/J=10$, $\gamma/J=0.1$, $J_{jj} = 1$ if $j$ is odd and 0 otherwise). (d), The same as in (c), but with added measurement of the magnetization at odd sites $\hat{D}_y=\hat{M}_\mathrm{odd}=\hat{N}_{\uparrow \text{odd}}-\hat{N}_{\downarrow \text{odd}}$. This protects the doubly occupied sites, thus, demonstrating protection of fermion pairs by measurement. The distribution of $N_{\mathrm{odd}}$ vanishes for odd numbers, implying that the fermions tunnel only in pairs. Simulations for 1D lattice.}
\end{figure}

For bosons (Fig.~\ref{panelfluc}a), the number fluctuations $\sigma^2_D$ calculated in the ground state decrease monotonically for increasing $U$, reflecting the superfluid - Mott insulator quantum phase transition. The measured state on the other hand behaves very differently and $\langle \sigma^2_D \rangle_{\mathrm{traj}}$ varies non-monotonically. For weak interaction, the fluctuations are strongly squeezed below those of the ground state; then they quickly increase, reach their maximum and subsequently decrease as the interaction becomes stronger. We explain this effect looking at the dynamics of single trajectories (cf. Fig.~\ref{panel2}). For small values of $U/J$, the population imbalance between odd and even sites oscillates and its uncertainty is squeezed by the measurement as described in Section III. However, when such oscillations reach maximal amplitude the local atomic repulsion spreads the atomic distribution and prevents the formation of states with large atom number in one of the two modes (Fig.~\ref{panel2}a). Since the interaction is a nonlinear term in the atomic dynamics, states with different imbalance oscillate  with different frequencies and the measurement is not able to squeeze the fluctuations of $\hat{N}_\text{odd}$ as efficiently as in the non-interacting case.
This behavior in the weak measurement and weak interaction limit is in contrast with what the effective two-sites model introduced in the previous section predicts (in a double well fluctuations are monotonically decreasing with $U$). However, the solution to this model is only valid in the $N \gg 1$ limit which suggests that this difference is due to the fact that in a double well with a large occupancy the population transfer is sequential, i.e.~atoms transfer one by one from one well to the other, increasing the total interaction energy, which goes up as $\langle\hat{n}_i^2\rangle,$ very steadily. On the other hand, in a lattice a collective excitation of a single atom from each site in one mode to another site in the other mode will increase the energy by $KU$, where $K$ is the number sites in one mode. This is an increase by a factor of $K$ compared to a single particle-hole excitation pair.  Therefore, a lattice and a global, long-range measurement scheme are necessary to observe such a collective transfer of atoms.

For weak measurement, but in the strongly interacting limit, we note that the measurement leads to a significant increase in fluctuations compared to the ground state. Both, the local interaction and measurement squeeze fluctuations, but as the measurement destroys the Mott insulator, the fluctuations are larger than in the ground state. Using first-order perturbation theory the ground state of the system is
\begin{align}
| \Psi_{J/U} \rangle = \left[ 1 + \frac{J}{U} \sum_{\langle i, j \rangle} b^\dagger_i b_j \right] | \Psi_{0} \rangle
\end{align}
where $| \Psi_{0}\rangle$ is the Mott insulator state and the second term represents a uniform distribution of particle-hole excitation pairs across the lattice. The action of a single photocount will amplify the present excitations increasing the fluctuations in the system. In fact, consecutive detections lead to an exponential growth of these excitations as for $K\gg1$ and unit filling, and the atomic state after $m$ quantum jumps becomes $\hat{c}^m| \Psi_{J/U} \rangle \propto | \Psi_{J/U} \rangle  +  | \Phi_m \rangle$ where 
\begin{align}
\label{eq:exc}
 | \Phi_m \rangle = \frac{2^mJ}{KU} \sum_{i \,\text{odd}} \left ( b^\dagger_i b_{i-1}  - b^\dagger_{i-1} b_i - b^\dagger_{i+1} b_i + b^\dagger_i b_{i+1} \right)| \Psi_0 \rangle.
\end{align}
In the weak measurement regime the effect of the non-Hermitian decay is negligible compared to the local atomic dynamics combined with the quantum jumps and so there is minimal dissipation occurring. Therefore, because of the exponential growth of the excitations, even a small number of photons arriving in succession can destroy the Mott insulator state very quickly. This will always happen given sufficient time, and provided there are finite fluctuations present in the initial state and so this will happen at any value of $U/J$, except for $J=0$ when the ground state becomes a single Fock state with no particle-hole excitations.

In the strong measurement regime ($\gamma \gg J$) the measurement becomes more significant than the local dynamics and the system will freeze the state in the measurement operator eigenstates. In this case, the squeezing will always be better than in the ground state, because measurement and on-site interaction cooperate in suppressing fluctuations. For low interaction strengths this should be obvious from the fact that in a superfluid ground state the atoms are spread out over the entire lattice and thus the uncertainty in atom number is large whereas measurement eigenstates have a well-defined occupation number. However, the strongly interacting regime is much less evident, especially since we have demonstrated how sensitive the Mott insulating state is to the quantum jumps when the measurement is weak.

To understand the strongly interacting case we will again use first-order perturbation theory and consider a postselected $\langle \hat{D}^\dagger \hat{D} \rangle = 0$ trajectory. This corresponds to a state that scatters no photons and so the non-Hermitian correction to the Hamiltonian is sufficient to understand the measurement. Since squeezing depends on the measurement strength and is common to all possible trajectories we can gain some insight by considering this specific case. However, we will now consider $\hat{D} = \Delta \hat{N} = \hat{N}_\text{odd} - \hat{N}_\text{even}$ as this measurement also has only two modes, $R = 2$, but its $\langle \hat{D}^\dagger \hat{D} \rangle = 0$ trajectory corresponds to the Mott insulating ground state. According to perturbation theory the modified ground state is now
\begin{align}
\label{perturbation}
| \Psi_{(J,U,\gamma)} \rangle = \left[ 1 + \frac{J}{U - i4\gamma} \sum_{\langle i, j \rangle} b^\dagger_i b_j \right] | \Psi_{0} \rangle.
\end{align}
The variance of the measurement operator for this state is given by
\begin{align}
\sigma^2_{\Delta N} = \frac{8 J^2 L}{U^2 + 16 \gamma^2} \nu(\nu+1),
\end{align}
where $\nu$ is the filling factor. From the form of the denominator we immediately see that both interaction and measurement squeeze with the same quadratic dependence and that the squeezing is always better than in the ground state (corresponding to $\gamma = 0$) regardless of the value of $U/J$. Also, depending on the ratio of $\gamma/U$ the squeezing can be dominated by measurement ($\gamma/U \gg 1$) or by interaction ($\gamma/U \ll 1$) or both processes can contribute equally ($\gamma/U \approx 1$). The $\hat{D} = \hat{N}_\text{odd}$ measurement should have a similar dependence on $\gamma$ and $U$ and be proportional to $(U^2 + \gamma^2)^{-1}$ since the $\gamma$ coefficient in the perturbative expansion depends on the value of $(J_{i,i} - J_{i-1,i-1})^2$. We can see the system transitioning into the strong measurement regime in Fig. \ref{panelfluc}(a) as the $U$-dependence flattens out with increasing measurement strength. In typical many-body systems in absence of measurement, strong correlations are a product of large interactions and the non-interacting limit reduces essentially to a single particle theory. In contrast, here, the measurement is another mechanism generating entanglement and strong correlations. This is more evident in weak interacting limit, where the measurement-induced interaction takes over the standard local interaction (as presented in the case of giant oscillations). We will also see in the next section how strong measurement leads to long-range correlated tunneling events.

For fermions (Figs.~\ref{panelfluc}b,c), the ground state of the attractive Hubbard model in the strong interacting regime contains mainly doubly occupied sites (pairs) and empty sites. Therefore, in absence of measurement, the fluctuations in the atom population $\hat{D}_x=\hat{N}_{\uparrow \text{odd}}+ \hat{N}_{\downarrow \text{odd}}$ ($\sigma^2_{D_{x}}$) increase with $U/J$ while the ones in the magnetization 
$\hat{D}_y=\hat{M}_\mathrm{odd}=\hat{N}_{\uparrow \text{odd}}-\hat{N}_{\downarrow \text{odd}}$  ($\sigma^2_{D_{y}}$) decrease because singly occupied sites become more improbable. The measurement induces two different kinds of dynamics using the same mode functions since, depending on the light polarization, we can address either the total population, or both total population and magnetization. In the first case (Figs.~\ref{panelfluc}b,c), the weak measurement quickly squeezes $\sigma^2_{D_{x}}$, but destroys the pairs as it does not distinguish between singly or doubly occupied sites. Figure~\ref{panel2}c illustrates such a measurement-induced breakup: the initial state contains mainly even values of $\hat{N}_\text{odd}$, corresponding to a superposition of empty and doubly occupied lattice sites. As time progresses and photons are scattered from the atoms, the variance $\sigma^2_{D_{x}}$ is squeezed and unpaired fermions are free to tunnel across the lattice allowing odd values for  $\hat{N}_\text{odd}$. In contrast, probing both density and magnetization reduces both their fluctuations and increase the lifetime of doubly occupied sites. The resulting measurement-induced dynamics is illustrated in Figure~\ref{panel2}(d): Atoms tunnel only in pairs with opposite spin as the probability distribution of $\hat{N}_\text{odd}$ contains only even values, hence demonstrating the measurement-based protection of fermion pairs. In both cases, the dynamics of the system is not a result of the projective quantum Zeno effect (here measurement is weak) but is a manifestation of the squeezing of the atomic population and magnetization of the two macroscopically occupied modes.

\begin{figure*}
\includegraphics[width=\textwidth]{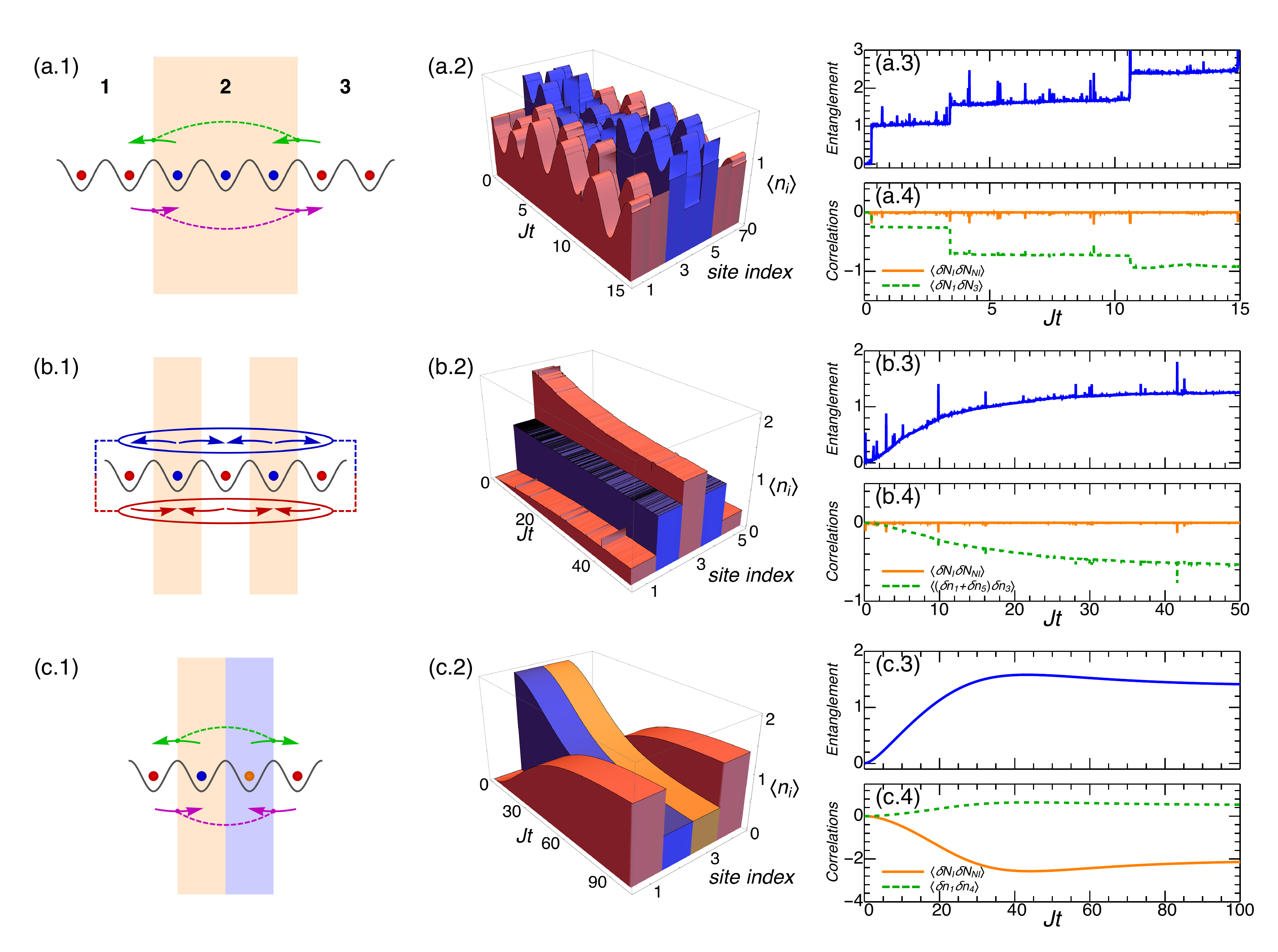}
\caption{\label{panelfreezing} Long-range correlated tunneling and entanglement, dynamically induced by strong global measurement in a single quantum trajectory. Panels (a),(b) and (c) show different measurement geometries, implying different constraints. Panels (1): Schematic representation of the long-range tunneling processes, when standard tunneling between different zones is Zeno-suppressed. Panels (2): Evolution of on-site densities; atoms effectively tunnel between disconnected regions due to correlations. Panels (3): Entanglement entropy growth between illuminated and non-illuminated regions. Panels (4): Correlations between different modes (solid orange line) and within the same mode (dashed green line); atom number $N_I$ ($N_{NI}$) in illuminated (non-illuminated) mode. 
(a) Atom number in the central region is frozen: The system is divided into three regions and correlated tunneling occurs between non-illuminated zones (a.1). Standard dynamics happens within each region, but not between them (a.2). Entanglement build up (a.3). Negative correlations between non-illuminated regions (dashed green line) and zero correlations between the $N_I$ and $N_{NI}$ modes (solid orange line) (a.4). Initial state: $|1,1,1,1,1,1,1 \rangle$, $\gamma/J=100$, $J_{jj}=[0,0,1,1,1,0,0]$.
(b) Even sites are illuminated, freezing $N_\text{even}$ and $N_\text{odd}$. Long-range tunneling is represented by any pair of one blue and one red arrow (b.1). Correlated tunneling occurs between non-neighboring sites without changing mode populations (b.2). Entanglement build up (b.3). Negative correlations  between edge sites (dashed green line) and zero correlations between the modes defined by $N_\text{even}$ and $N_\text{odd}$ (solid orange line) (b.4). Initial state: $|0,1,2,1,0 \rangle$,  $\gamma/J=100$, $J_{jj}=[0,1,0,1,0]$.
(c) Atom number difference between two central sites is frozen. Correlated tunneling leads to exchange of long-range atom pairs between illuminated and non-illuminated regions (c.1,2). Entanglement build up (c.3). In contrast to previous examples, sites in the same zones (illuminated/ non-illuminated) are positively correlated (dashed green line), while atoms in different zones are negatively correlated (solid orange line) (c.4). Initial state: $|0,2,2,0 \rangle$, $\gamma/J=100$, $J_{jj}=[0,-1,1,0]$. 1D lattice, $U/J=0$, the trajectory is smooth as no photons are detected.}
\end{figure*}

\section{Emergent long-range correlated tunneling} 
When $\gamma \gg J$, the photodetections freeze the modes' atom number, and decorrelate the populations of different modes. In the quantum Zeno limit of projective measurement ($\gamma \rightarrow \infty$) tunneling through the boundaries between modes would be fully suppressed. However, by considering a finite $\gamma/J$ we observe additional dynamics while the usual atomic tunneling is still strongly Zeno-suppressed. In this regime, we observe the following effects. First, the evolution between nearest neighbors within each mode is basically unperturbed by the measurement process and it is determined by the usual tunneling. Importantly, by engineering the light mode functions, it is possible to forbid part of this first-order dynamics and select which processes participate in the quantum Zeno dynamics. Therefore, one can design the Zeno subspace of the Hilbert space where the system evolves. Second, tunneling between different spatial modes is possible only via higher-order, long-range correlated tunneling events that preserve the eigenvalue of $\hat{D}$. In other words, atoms tunnel across distant sites of the lattice via a virtual state and they behave as delocalized correlated pairs. 

We can, once more, gain insight into this process from the non-Hermitian Hamiltonian. By looking at a second order expansion \cite{Auerbach,Kozlowski2015NH} of the Hamiltonian confined to a Zeno subspace of a two-mode measurement, $R=2$, for $U/J = 0$ we obtain the following effective Hamiltonian
\begin{align}
\label{eq:hz}
\hat{H}_Z = \hat{P}_0 \left[ -J \sum_{\langle i, j \rangle} b^\dagger_i b_j - i \frac{J^2} {A \gamma} \sum_{\varphi} 
\sum_{\substack{\langle i \in \varphi, j \in \varphi^\prime \rangle \\ \langle k \in \varphi^\prime, l \in \varphi \rangle}} b^\dagger_i b_j b^\dagger_k b_l \right] \hat{P}_0,
\end{align}
where $\hat{P}_0$ is the projector into the Zeno subspace, $A = (J_\varphi - J_{\varphi^\prime})^2$ is a constant that depends on the measurement scheme, $\varphi$ denotes a set of sites belonging to a single mode and $\varphi^\prime$ is the set's complement (e.g. odd and even sites).

We can infer a lot of information from this simple expression. First, we see that first order tunneling will only survive between neighboring sites that belong to the same mode, because $\hat{P}_0 b^\dagger_i b_j \hat{P}_0 = 0$ otherwise. Second, the second-order term exists between different modes and occurs at a rate $\sim J^2/\gamma$. The imaginary prefactor means that this tunneling behaves like an exponential decay (overdamped oscillations). This picture is consistent with the projective limit ($\gamma \rightarrow \infty$), where the atomic state is constrained to an eigenspace of the measurement operator.

Crucially, what sets this effect apart from usual many-body dynamics with short-range interactions is that first order processes are selectively suppressed by the global conservation of the measured observable and not by the prohibitive energy costs of doubly-occupied sites, as is the case in the $t-J$ model \cite{Auerbach}. This has profound consequences as this is the physical origin of the long-range correlated tunneling events represented in \eqref{eq:hz} by the fact that sites $j$ and $k$ can be very distant. This is because the projection $\hat{P}_0$ is not sensitive to individual site occupancies, but instead enforces a fixed value of the observable, i.e.~a single Zeno subspace.

Illuminating only the central region of the optical lattice and detecting light in the diffraction maximum, we freeze the atom number $\hat{N}_\text{illum}$~\cite{MekhovPRL2009,MekhovPRA2009} (Fig.~\ref{panelfreezing}a). The measurement scheme defines two different spatial modes: the non-illuminated zones $1$ and $3$ and the illuminated one~$2$. Figure~\ref{panelfreezing}a.2 illustrates the evolution of the mean density at each lattice site: typical dynamics occurs within each region but the standard tunneling between different zones is suppressed. Importantly, processes that do not change $N_\text{illum}$ are still possible since an atom from $1$ can tunnel to $2$, if simultaneously one atom tunnels from $2$ to $3$. Thus, effective long-range tunneling between two spatially disconnected zones $1$ and $3$ happens due to two-step processes $1\rightarrow2\rightarrow3$ or $3\rightarrow2\rightarrow1$. These transitions are responsible for the negative (anti-)correlations $\langle \delta N_1 \delta N_3\rangle = \langle N_1 N_3\rangle-\langle N_1 \rangle\langle N_3\rangle$ showing that an atom disappearing in 1 appears in 3, while there are no number correlations between illuminated and non-illuminated regions,  $\langle( \delta N_1 +\delta N_3)\delta N_2\rangle =0$ (Fig.~\ref{panelfreezing}a.4). In contrast to fully-projective measurement, the intermediate (virtual) step in the correlated tunneling process builds long-range entanglement between illuminated and non-illuminated regions (Fig.~\ref{panelfreezing}a.3).

To make correlated tunneling visible even in the mean atom number, we suppress the standard Bose-Hubbard dynamics by illuminating only the even sites of the lattice~(Fig.~\ref{panelfreezing}b). Even if this measurement scheme freezes both $N_\text{even}$ and $N_\text{odd}$, atoms can slowly tunnel between the odd sites of the lattice, despite them being spatially disconnected. This atom exchange spreads correlations between non-neighboring lattice sites on a time scale $\sim \gamma/J^2$. The schematic explanation of long-range correlated tunneling is presented in Fig.~\ref{panelfreezing}b.1: the atoms can tunnel only in pairs to assure the globally conserved values of $N_\text{even}$ and $N_\text{odd}$, such that one correlated tunneling event is represented by a pair of one red and one blue arrow. Importantly, this scheme is fully applicable for a lattice with large atom and site numbers, well beyond the numerical example in Fig.~\ref{panelfreezing}b.1, because  as we can see in \eqref{eq:hz} it is the geometry of quantum measurement that assures this mode structure (in this example, two modes at odd and even sites) and therefore underlying pairwise global tunneling. Such long-range correlations and long-range entanglement (even in a 1D system) can develop essentially because of the coherent global addressing. In contrast, the local uncorrelated probing of individual sites would decreases the probability of individual tunneling events being correlated. 

This global pair tunneling may play a role of a building block for more complicated many-body effects. For example, a pair tunneling between the neighbouring sites has been recently shown to play important role in the formation of new quantum phases, e.g., pair superfluid \cite{LewensteinPSFNJP} and lead to formulation of extended Bose-Hubbard models \cite{LewensteinExtBHM}. The search for novel mechanisms providing long-range interactions is crucial in many-body physics. One of the standard candidates is the dipole-dipole interaction in, e.g., dipolar molecules, where the mentioned pair tunneling between even neighboring sites is already considered to be long-range \cite{LewensteinPSFNJP,LewensteinExtBHM}. In this context, our work suggests a fundamentally different mechanism originating from quantum optics: the backaction of global and spatially structured measurement, which as we prove can successfully compete with other short-range processes in many-body systems. This opens promising opportunities for future research.

The scheme in Fig.~\ref{panelfreezing}b.1 can help to design a nonlocal reservoir for the tunneling (or ``decay'') of atoms from one region to another. For example, if the atoms are placed only at odd sites,  according to \eqref{eq:hz} their tunnelling is suppressed since the multi-tunneling event must be successive, i.e.~an atom tunnelling into a different mode, $\varphi^\prime$, must then also tunnel back into its original mode, $\varphi$. If, however, one adds some atoms to even sites (even if they are far from the initial atoms), the correlated tunneling events become allowed and their rate can be tuned by the number of added atoms. This resembles the repulsively bound pairs created by local interactions \cite{Winkler2006,BlochNature2007}. In contrast, here the atom pairs are long-range correlated due to the global measurement. Additionally, these long-range correlations are a consequence of the dynamics being constrained to a Zeno subspace: the virtual processes allowed by the measurement entangle the spatial modes nonlocally. Since the measurement only reveals the total number of atoms in the illuminated sites, but not their exact distribution, these multi-tunelling events cause the build-up of long range entanglement. This is in striking contrast to the entanglement caused by local processes which can be very confined, especially in 1D where it is typically short range. This makes numerical calculations of our system for large atom numbers really difficult, since well-known methods such as Density Matrix Renormalization Group and Matrix Product States \cite{Schollwock} (which are successful for short-range interactions) rely on the limited extent of entanglement.

These types of configurations open intriguing opportunities for quantum engineering of system - bath interactions: here both the system and reservoir, represented by different modes, are many-body systems with internal long-range entanglement.

The negative number correlations are typical for systems with constraints (superselection rules) such as fixed atom number. The effective dynamics due to our global, but spatially structured, measurement introduces more general constraints to the evolution of the system. For example, in Fig.~\ref{panelfreezing}c we show the generation of positive number correlations (shown in Fig.~\ref{panelfreezing}c.4) by freezing the atom number difference between the sites ($N_\text{odd}-N_\text{even}$, by measuring at the diffraction minimum). Thus, atoms can only enter or leave this region in pairs, which again is possible due to correlated tunneling (Figs.~\ref{panelfreezing}c1-2) and manifests positive correlations. Since this corresponds to a no photon trajectory, we can solve exactly for a small system and for two atoms, these correlations grow as $\langle \delta \hat{n}_1 \delta \hat{n}_4 \rangle \approx [1 - \sech^2(4J^2t/\gamma)]/4$. As in the previous example, two edge modes in Fig.~\ref{panelfreezing}c can be considered as a nonlocal reservoir for two central sites, where a constraint is applied. Note that, using more modes, the design of higher-order multi-tunneling events is possible.

\section{Summary}
We proved that the quantum backaction of a global measurement can efficiently compete with standard local processes in strongly correlated systems. This introduces a physically novel source of competition in research on quantum many-body systems. The competition becomes efficient due to the ability to spatially structure the global measurement at a microscopic scale comparable to the lattice period, without the need for single site addressing. The extreme tunability of the setup we considered allows us to vary the spatial profile of the measurement operator, effectively tailoring the long-range entanglement and long-range correlations present in the system. The competition between the global backaction and usual atomic dynamics leads to the production of spatially multimode macroscopic superpositions which exhibit large-scale oscillatory dynamics and could be used for quantum information and metrology. Such dynamical states show spatial density-density correlations with nontrivial periods and long-range coherence, thus having supersolid properties, but as an essentially dynamical version. We showed the possibility of measurement-induced break-up and protection of strongly interacting fermion pairs. In the strong measurement regime, the usual nearest-neighbour tunnelling is suppressed but the atoms can still tunnel across the lattice because of correlated tunnelling. Such globally paired tunneling due to a fundamentally novel phenomenon can enrich physics of long-range correlated systems beyond relatively short-range interactions expected from standard dipole-dipole interactions \cite{LewensteinPSFNJP,LewensteinExtBHM}. These nonlocal high-order processes entangle regions of the optical lattice that are disconnected by the measurement. Using different detection schemes, we showed how to tailor density-density correlations between distant lattice sites. Quantum optical engineering of nonlocal coupling to environment, combined with quantum measurement, can allow the design of nontrivial system-bath interactions, enabling new links to quantum simulations~\cite{Stannigel2013,Elliott2015b} and thermodynamics~\cite{Erez2008} and extend these directions to the field of non-Hermitian quantum mechanics, where quantum optical setups are particularly promising~\cite{LeePRL2014}. Importantly, both systems and baths, designed by our method, can be strongly correlated systems with internal long-range entanglement.

Our predictions can be tested using both macroscopic measurements~\cite{EsslingerNat2010,HemmerichScience2012,ZimmermannPRL2014,KetterlePRL2011} as well as novel methods based on single-site resolution~\cite{BlochNature2011,Weitenberg2011,GreinerNature2009,patil2014}. A pathway to realize this is to combine several recent experimental breakthroughs: a BEC was trapped in a cavity, but without a lattice~\cite{EsslingerNat2010,HemmerichScience2012,ZimmermannPRL2014}; detection of light scattered from truly ultracold atoms in optical lattices was performed, but without a cavity~\cite{Weitenberg2011,KetterlePRL2011}, and very recently an optical lattice has been obtained in a cavity \cite{Landig2015a, Klinder2015}. Furthermore, the single-atom and multi-particle quantum Zeno effect was observed by light scattering~\cite{patil2014,Barontini2015}.

A source of decoherence affecting an experimental realisation of our setup is photon miscounts due to imperfect detectors. This is more important in the weak measurement regime but its effect strongly depends on the detection scheme (multicomponent states are more fragile than single component ones). For strong measurement, the exact number of photons detected is not important as the state of the system is determined by the photon emission rate. Atom losses caused by spontaneous emission and heating are another source of error that increase the uncertainty of the atomic state. Based on off-resonant scattering and thus being non-sensitive to a detailed level structure, our approach can be applied to other arrays of natural or artificial quantum objects: molecules (including biological ones)~\cite{MekhovLP2013}, ions \cite{Blatt2012}, atoms in multiple cavities~\cite{Hartmann2006}, semiconductor~\cite{Trauzettel2007} or superconducting~\cite{Fink2009} qubits.

\begin{acknowledgments}
The work was supported by the EPSRC (DTA and EP/I004394/1).

G. M. and W. K. contributed equally to this work. 
\end{acknowledgments}

\section*{Appendix A: Atomic Hamiltonians and light-matter coupling}
We describe atomic dynamics by the (Bose-) Hubbard Hamiltonian. For the bosonic case,
\begin{align}\label{hamiltonianB}
\hat{H}_0=-\hbar J \sum_{\langle i,j\rangle}b_j^\dagger b_i +  \frac{\hbar U}{2} \sum_i \hat{n}_i \left(\hat{n}_i-1\right),
\end{align}
while for the fermionic case
\begin{align}\label{hamiltonian}
\hat{H}_0=-\hbar J \sum_{\sigma=\uparrow,\downarrow} \sum_{\langle i,j\rangle} f_{j,\sigma}^\dagger f_{i,\sigma} -\hbar U \sum_i \hat{n}_{i,\uparrow}\hat{n}_{i,\downarrow},
\end{align}
where $b$ and $f_\sigma$ are respectively the bosonic and fermionic annihilation operators, $\hat{n}$ is the atom number operator, and $U$ and $J$ are the on-site interaction and tunneling coefficients.
The Hamiltonian of the light-matter system is~\cite{Mekhov2012}
\begin{align}\label{fullH}
\hat{H}=\hat{H}_0+ \sum_l \hbar \omega_l a_l^\dagger a_l + \hbar \sum_{l,m}\Omega_{lm}a_l^\dagger a_m \hat{F}_{lm}
\end{align}
where $a_l$ are the photon annihilation operators for the light modes with frequencies $\omega_l$, $\Omega_{lm}=g_l g_m/\Delta_a$, $g_l$ are the atom-light coupling constants, and $\Delta_a=\omega_p-\omega_a$ is the probe-atom detuning. The operator $\hat{F}_{lm}=\hat{D}_{lm}+\hat{B}_{lm}$ couples atomic operators to the light fields:
\begin{align}
&\hat{D}_{lm}=\sum_{i} J^{lm}_{ii} \hat{n}_i, \quad \hat{B}_{lm}=\sum_{\langle i,j \rangle} J^{lm}_{ij} b_i^{\dagger} b_j, \\
&J^{lm}_{ij}=\int \! w (\b{r} -\b{r}_i) u_{l}^*(\b{r})u_{m}(\b{r}) w (\b{r} -\b{r}_j) \, \mathrm{d}  \b{r}.
\end{align}
$\hat{F}_{lm}$ originates from the overlaps between the light mode functions $u_l({\bf r})$ and density operator $\hat{n}({\bf r})=\hat{\Psi}^\dag({\bf r})\hat{\Psi}({\bf
  r})$, after the matter-field operator is expressed via Wannier
functions: $\hat{\Psi}({\bf r})=\sum_i b_i w({\bf r}-{\bf
  r}_i)$. $\hat{D}_{lm}$ sums the density contributions $\hat{n}_i$,
while $\hat{B}_{lm}$ sums the matter-field interference
terms. The light-atom
coupling via operators assures the dependence of light on the atomic
quantum state. These equations can be extended to fermionic atoms introducing an additional index for the polarization of light modes.

From (\ref{hamiltonian}) we can compute the Heisenberg equation for light operators in the stationary limit~\cite{Mekhov2012}. Specifically, we consider two light modes: a coherent probe beam $a_0$ and the scattered light $a_1$, which is enhanced by a cavity with decay rate $\kappa$.  This allows us to express the cavity mode annihilation operator as $a_1 =C\hat{F}_{10}$, where
\begin{align}\label{defC}
C=\frac{i \Omega_{10} a_0}{i \Delta_p - \kappa},
\end{align}
$\Delta_p=\omega_0-\omega_1$ is the probe-cavity detuning, and $C$ is the cavity analog of the Rayleigh scattering coefficient in free space~\cite{Kozlowski}. In the main text, we drop the subscript in $a_1$ and superscripts in $J^{lm}_{ij}$.

\section*{Appendix B: Modeling the measurement process}
Quantum trajectories describe the evolution of a quantum system conditioned on the result of a measurement~\cite{MeasurementControl}. It is possible to represent dynamics of a system with two different processes: non-Hermitian dynamics and quantum jumps. The first one describes evolution of the system between two consecutive measurement events, while the second one models the photodetections. These can be simulated introducing the jump operator $c=\sqrt{2 \kappa} a_1$ and effective Hamiltonian $\hat{H}_{\mathrm{eff}}=\hat{H}_0- i \hbar c^\dagger c/2$. Starting from the time $t_0$, a random number $r$ between 0 and 1 is generated with uniform probability, and we solve the effective Schr\"odinger equation
\begin{align}
\frac{\mathrm{d}}{\mathrm{d}t} | \psi(t) \rangle = - \frac{i}{\hbar}  \hat{H}_{\mathrm{eff}} | \psi(t) \rangle
\end{align}
until the time moment $t_j$ such that $\langle \psi(t_j) | \psi(t_j) \rangle =r$. At this time moment, a photon is detected and the jump operator is applied to the state of the system, which is subsequently normalized:
\begin{align}
 | \psi(t_j) \rangle \rightarrow \frac{c| \psi(t_j) \rangle}{\sqrt{\langle \psi(t_j) |c^\dagger c| \psi(t_j) \rangle}}.
\end{align}
Finally, a new random number is generated and the procedure is iterated setting $t_j$ as a new starting time. This technique allows us to simulate a single quantum trajectory, thus modeling the result of a single experimental run.

\bibliographystyle{apsrev4-1}
\bibliography{references}

\end{document}